# Knowledge-fused differential dependency network models for detecting significant rewiring in biological networks

Ye Tian[1,†], Bai Zhang[2,†], Eric P. Hoffman[3], Robert Clarke[4], Zhen Zhang[2], Ie-Ming Shih[2], Jianhua Xuan[1], David M. Herrington[5], and Yue Wang[1,*]

[1]Department of Electrical & Computer Engineering, Virginia Tech, Arlington, VA 22203, USA; [2]Department of Pathology, Johns Hopkins University, Baltimore, MD 21231, USA; [3]Research Center for Genetic Medicine, Children's National Medical Center, Washington, DC 20010, USA; [4]Lombardi Comprehensive Cancer Center, Georgetown University, Washington, DC 20057, USA; [5]Department of Medicine, Wake Forest University, Winston-Salem, NC 27157, USA

## ABSTRACT

**Motivation:** Modeling biological networks serves as both a major goal and an effective tool of systems biology in studying mechanisms that orchestrate the activities of gene products in cells. Biological networks are context specific and dynamic in nature. To systematically characterize the selectively activated regulatory components and mechanisms, the modeling tools must be able to effectively distinguish significant rewiring from random background fluctuations. While differential networks cannot be constructed by existing knowledge alone, novel incorporation of prior knowledge into data-driven approaches can improve the robustness and biological relevance of network inference. However, the major unresolved roadblocks include: big solution space but a small sample size; highly complex networks; imperfect prior knowledge; missing significance assessment; and heuristic structural parameter learning.

**Results:** To address these challenges, we formulated the inference of differential dependency networks that incorporates both conditional data and prior knowledge as a convex optimization problem, and developed an efficient learning algorithm to jointly infer the conserved biological network and the significant rewiring across different conditions. We used a novel sampling scheme to estimate the expected error rate due to “random” knowledge and based on which, developed a strategy that fully exploits the benefit of this data-knowledge integrated approach. We demonstrated and validated the principle and performance of our method using synthetic datasets. We then applied our method to yeast cell line and breast cancer microarray data and obtained biologically plausible results.

**Availability:** The open-source R software package is freely available at http://www.cbil.ece.vt.edu/software.htm.

## 1 INTRODUCTION

Biological networks are context-specific and dynamic in nature (Mitra, et al., 2013). Under different conditions, different regulatory components and mechanisms are selectively activated or deactivated (Califano, 2011; Creixell, et al., 2012). For example, the topology of underlying biological network changes in response to internal or external stimuli, where cellular components exert their functions through interactions with other molecular components (Barabasi, et al., 2011; Ideker and Krogan, 2012). Thus, in addition to asking “which genes are differentially expressed”, the new question is “which genes are differentially connected?” (Hudson, et al., 2012; Reverter, et al., 2010). Studies on network-attacking events will shed new light on whether network rewiring is a general principle of biological systems regarding disease progression or therapeutic responses (Califano, 2011; Creixell, et al., 2012). Moreover, due to inevitable experimental noise, snapshot of dynamic expression, and post-transcriptional or translational/post-translational modifications, systematic efforts to characterize biological networks must effectively distinguish significant network rewiring from random background fluctuations (Mitra, et al., 2013).

Almost exclusively using high throughput gene expression data and focusing on conserved biological networks, various network inference approaches have been proposed and tested (Mitra, et al., 2013), including probabilistic Boolean networks (Shmulevich, et al., 2002), state-space models (Rangel, et al., 2004; Tyson, et al., 2011), and probabilistic graphical models (Friedman, 2004). However, since biological networks are assembled in single experimental conditions, they overlook the inherently dynamic nature of molecular interactions, which can be extensively rewired across different conditions. Hence, current network models reveal only partial snapshots of the cell. To explicitly address differential network analysis (Califano, 2011; Hudson, et al., 2009; Ideker and Krogan, 2012), some initial efforts have been recently reported (Mitra, et al., 2013). In our previous work, Zhang *et al*. proposed to model differential dependency networks between two conditions by detecting network rewiring using significance tests on local dependencies across conditions (Zhang, et al., 2009; Zhang, et al., 2011), which is a substantially different method from the one proposed in this paper where experimental data and prior knowledge are jointly modeled. The approach was successfully extended by Roy *et al*. to learning dynamic networks across multiple conditions (Roy, et al., 2011), and by Gill *et al*. to assessing the overall evidence of network differences between two conditions using the connectivity scores associated with a gene or module (Gill, et al., 2010). The time evolvement of network structures is examined with a fused penalty term to encode relationship between adjacent time points in (Ahmed and Xing, 2009). Furthermore, recent efforts have also been made to incorporate existing knowledge on network biology into data-driven network inference (Kanehisa and Goto, 2000). Wang *et al.* proposed to incorporate prior knowledge into the inference of conserved networks in a single condition by adjusting the Lasso penalties (Wang, et al., 2013). Yet, the inher-ently dynamic wiring of biological networks remains under-explored at the systems level, as interaction data are typically reported under diverse while isolated conditions (Mitra, et al., 2013).

* To whom correspondence should be addressed.

† These authors contributed equally.

There are at least five unsolved issues concerning differential network inference using data-knowledge integrated approaches: (1) the solution (search) space is usually large while sample sizes are small, resulting in potential overfitting; (2) both conserved and differential biological networks are complex while currently lacking closed-form or efficient numerical solutions; (3) "structural" model parameters are assigned heuristically, leading to potentially suboptimal solutions; (4) prior knowledge is imperfect for inferring biological networks under specific conditions, *e.g.*, false positive "connections", biases, and non-specificity; and (5) most current methods do not provide significance assessment on the differential connections and rigorous testing of the type I error rate.

To address these challenges, we formulate the inference of differential dependency networks that incorporates both conditional data and prior knowledge as a convex optimization problem, and develop an efficient learning algorithm to jointly infer the conserved biological network and the significant rewiring across different conditions. Extending and improving our work on Gaussian graphical models (Tian, et al., 2011; Zhang and Wang, 2010), we design block-wise separable penalties in the Lasso-type models that permit joint learning and knowledge incorporation with an efficient closed-form solution. We estimate the expected error rate due to "random" prior knowledge via a novel sampling scheme and based on which, develop a strategy to fully exploit the benefit of this data-knowledge integrated approach. We determine the values of model parameters that quantitatively correspond to the expected significance level, and evaluate the statistical significance of each of the detected differential connections. We validate our method using synthetic datasets and comprehensive comparisons. We then apply our method to yeast cell line and breast cancer microarray data and obtain biologically plausible results.

## 2 METHODS

### 2.1 Formulation of knowledge-fused differential dependency network (kDDN)

We represent the condition-specific biological networks as graphs. Suppose there are $p$ nodes (genes) in the network of interest, and we denote the vertex set as $V$. Let $G^{(1)}=\left(V,E^{(1)}\right)$ and $G^{(2)}=(V,E^{(2)})$ be the two undirected graphs under the two conditions. $G^{(1)}$ and $G^{(2)}$ have the same vertex set $V$, and condition-specific edge sets $E^{(1)}$ and $E^{(2)}$. The edge changes indicated by the differences between $E^{(1)}$ and $E^{(2)}$ are of particular interest, since such rewiring may reveal pivotal information on how the organisms respond to different conditions. We label the edges as common edges or specific to particular condition in graph $G=(V,E)$ to represent the learned networks under the two conditions.

Prior knowledge on biological networks is obtained from well-established databases such as KEGG (Kanehisa and Goto, 2000) and is represented as a knowledge graph $G_{\mathbf{W}}=(V,E_{\mathbf{W}})$, where the vertex set $V$ is the same set of nodes (genes) and edge set $E_{\mathbf{W}}$ over $V$ is translated from prior knowledge. The adjacency matrix of $G_{\mathbf{W}}$, $\mathbf{W}\in\Re^{p\times p}$, is used to encode the prior knowledge. The elements of $\mathbf{W}$ are either 1 or 0, with $W_{ji}=1$ indicating the existence of an edge from the $j^{\text{th}}$ gene to the $i^{\text{th}}$ gene (or their gene products), where $i,j=1,2,\cdots,p,i\neq j$. $\mathbf{W}$ is symmetric if the prior knowledge is not directed.

The main task in this paper is to infer from data and prior knowledge $G_{\mathbf{W}}$ the condition-specific edge sets $E$, corresponding to $E^{(1)}$ and $E^{(2)}$. The method is illustrated in Figure 1..

We consider the $p$ nodes in $V$ as $p$ random variables, and denote them as $X_1,X_2,\cdots,X_p$. Suppose there are $N_1$ samples under condition 1 and $N_2$ samples under condition 2. Without loss of generality, we assume $N_1=N_2=N$. Under the first condition, for variable $X_i$, we have observations $\mathbf{x}_i^{(1)}=[x_{1i}^{(1)},x_{2i}^{(1)},\cdots,x_{Ni}^{(1)}]^T$, $i=1,2,\cdots,p$, while under the second condition, we have $\mathbf{x}_i^{(2)}=[x_{1i}^{(2)},x_{2i}^{(2)},\cdots,x_{Ni}^{(2)}]^T$, $i=1,2,\cdots,p$. Further, let $\mathbf{X}^{(1)}=[\mathbf{x}_1^{(1)},\mathbf{x}_2^{(1)},\cdots,\mathbf{x}_p^{(1)}]$ be the data matrix under condition 1 and $\mathbf{X}^{(2)}=[\mathbf{x}_1^{(2)},\mathbf{x}_2^{(2)},\cdots,\mathbf{x}_p^{(2)}]$ be the data matrix under condition 2.

Denote

$$\mathbf{y}_i=\begin{bmatrix}\mathbf{x}_i^{(1)}\\ \mathbf{x}_i^{(2)}\end{bmatrix},\quad \mathbf{X}=\begin{bmatrix}\mathbf{X}^{(1)} & \mathbf{0}\\ \mathbf{0} & \mathbf{X}^{(2)}\end{bmatrix}, \tag{1}$$

and

$$\boldsymbol{\beta}_i=\begin{bmatrix}\boldsymbol{\beta}_i^{(1)}\\ \boldsymbol{\beta}_i^{(2)}\end{bmatrix}=[\beta_{1i}^{(1)},\beta_{2i}^{(1)},\cdots,\beta_{pi}^{(1)},\beta_{1i}^{(2)},\beta_{2i}^{(2)},\cdots,\beta_{pi}^{(2)}]^T, \tag{2}$$

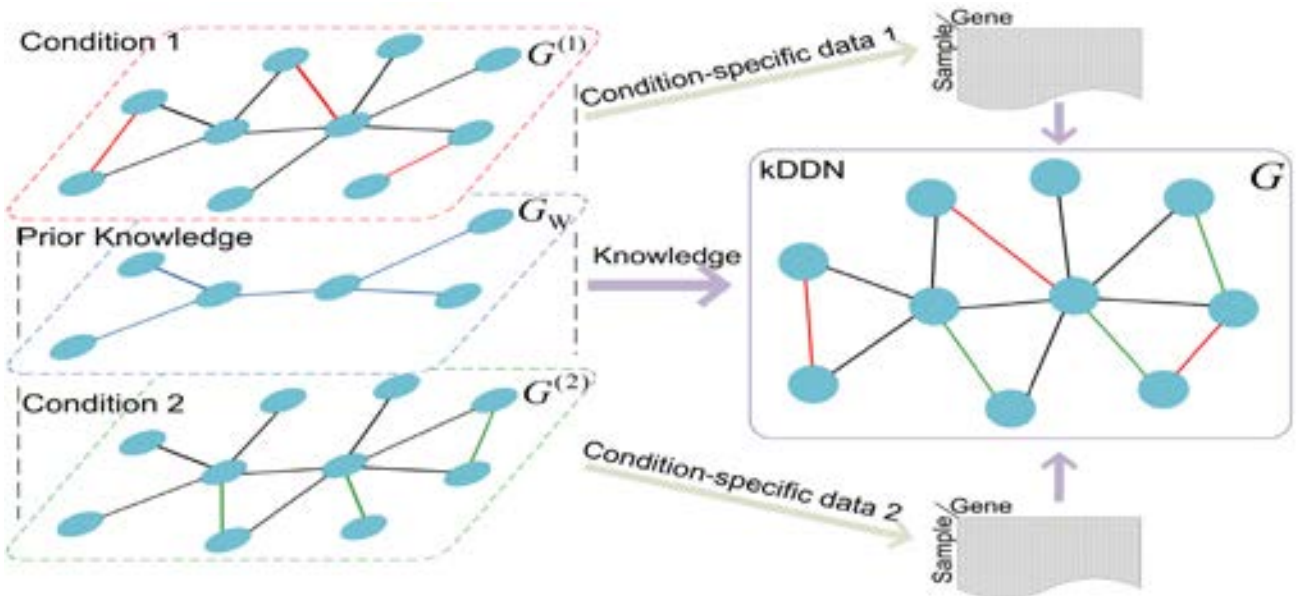

**Figure 1.** Knowledge-guided differential dependency network learning. Black edges are common edges. Red and green edges are differential edges specific to each condition.

with the non-zero elements of $\boldsymbol{\beta}_i^{(1)}$ indicate the neighbors of the $i^{\text{th}}$ node under the first condition and the non-zero elements of $\boldsymbol{\beta}_i^{(2)}$ indicate the neighbors of the $i^{\text{th}}$ node under the second condition.

The problem of simultaneously learning network structures under two conditions and their changes is formulated into a convex optimization problem with sparsity constraints. For each node (variable) $X_i$, $i=1,2,\cdots,p$, solve the optimization with the objective function

$$\begin{aligned} f(\boldsymbol{\beta}_i)=&\frac{1}{2}\left\|\mathbf{y}_i-\mathbf{X}\boldsymbol{\beta}_i\right\|_2^2+\lambda_1\sum_{j=1}^{p}\left(1-W_{ji}\theta\right)\left(|\beta_{ji}^{(1)}|+|\beta_{ji}^{(2)}|\right)\\ &+\lambda_2\left\|\boldsymbol{\beta}_i^{(1)}-\boldsymbol{\beta}_i^{(2)}\right\|_1. \end{aligned} \tag{3}$$

The non-zero elements in $\mathbf{W}$ introduce knowledge to the objective function (3), and $\theta$ is a $\ell_1$ penalty relaxation parameter taking value in $[0,1]$.

The solution is obtained by minimizing (3),

$$\begin{aligned} \boldsymbol{\beta}_i&=\arg\min_{\boldsymbol{\beta}_i} f(\boldsymbol{\beta}_i)\\ &=\arg\min_{\boldsymbol{\beta}_i^{(1)},\boldsymbol{\beta}_i^{(2)}}\frac{1}{2}\|\mathbf{y}_i-\mathbf{X}\boldsymbol{\beta}_i\|_2^2\\ &\quad+\lambda_1\sum_{j=1}^{p}(1-W_{ji}\theta)(|\beta_{ji}^{(1)}|+|\beta_{ji}^{(2)}|)+\lambda_2\|\boldsymbol{\beta}_i^{(1)}-\boldsymbol{\beta}_i^{(2)}\|_1\\ \text{s.t. } &\beta_{ii}^{(1)}=0,\beta_{ii}^{(2)}=0. \end{aligned} \tag{4}$$

The structures of the graphical model under two conditions are obtained jointly by solving (4) sequentially for all nodes. The inconsistency between $\boldsymbol{\beta}_i^{(1)}$ and $\boldsymbol{\beta}_i^{(2)}$ highlights the structural changes between two conditions, and the collection of differential edges form the differential dependency network.

Given the vast search space and complexity in both conserved and differential networks, it is crucial for kDDN to identify statistically significant network changes and filter the structural and parametric inconsistencies due to noise in the data and limited samples. The objective is achieved by selecting the proper model specified by $\lambda_1$ and $\lambda_2$ that best fits the data and suffices the statistical significance. $\lambda_1$ is determined by controlling the false discovery rate of edges and $\lambda_2$ is found by setting differential edges to defined significance level. We refer readers to Supplementary Information for detailed discussion of model parameter setting approaches.

With parameters specified, problem (4) can be solved efficiently by the block coordinate descent algorithm presented in Supplementary Information.

## 2.2 Incorporation of prior knowledge

The prior knowledge is explicitly incorporated into the formulation by $W_{ji}$ and $\theta$ in the block-wise weighted $\ell_1$-regularization term. $W_{ji}=1$ indicates that the prior knowledge supports an edge from the $j^{\text{th}}$ gene to the $i^{\text{th}}$ gene and 0 otherwise. A proper $\theta$ will reduce the penalty applied to $\beta_{ji}^{(c)}$ corresponding to the connection between $X_j$ and $X_i$ with $W_{ji}=1$. As a result, the connection between $X_j$ and $X_i$ will more likely be detected.

$\theta$ determines to what degree the knowledge will affect the inference. When $\theta=0$, the algorithm ignores all knowledge information and gives solely data based results; conversely, when $\theta=1$, the edge between $X_j$ and $X_i$ will always be included if such an edge exists in the prior knowledge. Thus, the prior knowledge will have a determining effect on the network inference and the equivalent modification of the formulation (4) may also affect other edges. Therefore the prior knowledge incorporation needs to find a proper balance between the experimental data and prior knowledge to achieve effective incorporation as well as limit the adverse effects caused by any spurious edges contained in imperfect prior knowledge.

Here we propose a strategy to control the adverse effects incurred in the worst-case scenario under which the given prior knowledge is totally random. In this case, the entropy of the knowledge distribution over the edges is maximized and the information introduced to the inference is minimal. Incorporating such random knowledge, the inference results will deviate from the purely data driven result. Then, $\theta$ is carefully chosen according to Theorem 1 so that the expected deviation is controlled within an acceptable range in the worst-case scenario. This approach guarantees the robustness even when the prior knowledge is highly inconsistent with the underlying ground-truth.

To quantify the effects of prior knowledge incorporation, we use Hamming distance between two adjacency matrices as a measurement for the dissimilarity of two graphs. Let $G_{\mathbf{T}}=(V,E_{\mathbf{T}})$ denote the ground-truth graph with edge set $E_{\mathbf{T}}$, $G_{\mathbf{X}}=(V,E_{\mathbf{X}})$ denote the graph learned purely from data, *i.e.* $\mathbf{W}=\mathbf{0}$, and $G_{\mathbf{X},\mathbf{W}_R,\theta}(V,E_{\mathbf{X},\mathbf{W}_R,\theta})$ denote the graph learned with prior knowledge. $\mathbf{W}_R$ indicates that the prior knowledge is "random". Let $d(G_*,G_\star)$ denote the Hamming distance between two graphs. Further, let $|E_*|$ be the number of edges in the graph $G_*$.

Our objective is to bound the increase of inference error rate associated with the purely data result, $d(G_{\mathbf{T}},G_{\mathbf{X},\mathbf{W}_R,\theta})/|E_{\mathbf{T}}|-d(G_{\mathbf{T}},G_{\mathbf{X}})/|E_{\mathbf{T}}|$, within an acceptable range $\delta$ even if the prior knowledge is the worst case by finding a proper $\theta$.

Since $G_{\mathbf{T}}$ is unknown, we instead control the increase in the error rate indirectly by evaluating the effect of random knowledge against $G_{\mathbf{X}}$, the purely data-driven inference result. Specifically, we use a sampling-based algorithm to find the empirical distribution of $d(G_{\mathbf{X}},G_{\mathbf{X},\mathbf{W}_R,\theta})$, and choose the largest $\theta\in[0,1]$ that satisfies:

$$\hat{\theta}=\max\theta \quad \text{s.t.} \quad \mathbb{E}[d(G_{\mathbf{X}},G_{\mathbf{X},\mathbf{W}_R,\theta})]/|E_{\mathbf{X}}|\le\delta. \tag{5}$$

A natural question is whether using $G_{\mathbf{X}}$ instead of $G_{\mathbf{T}}$ to control the increase in the error rate induced by random knowledge is legitimate. To answer this question, we show in Theorem 1 (proof included in Supplementary Information) that the $\theta$ obtained in (5) in fact controls an upper bound of $\mathbb{E}[d(G_{\mathbf{T}},G_{\mathbf{X},\mathbf{W_R},\boldsymbol{\theta}})]/|E_{\mathbf{T}}|$, *i.e.* the increase in the network inference error rate induced by random prior knowledge (the worst-case scenario), under the assumption that the number of false negatives ($FN$) in data-driven result $G_{\mathbf{X}}$ is smaller than the number of false positives ($FP$). As we adopt a strategy to refrain from falsely joining unconnected edges (Meinshausen and Bühlmann, 2006), this assumption generally holds.

**Theorem 1.** *For a given $\delta\in[0,1)$, if the prior knowledge incorporation parameter $\theta$ satisfies the inequality*

$$\frac{\mathbb{E}[d(G_{\mathbf{X}},G_{\mathbf{X},\mathbf{W}_R,\theta})]}{|E_{\mathbf{X}}|}\le\delta, \tag{6}$$

*then the increase in the error rate induced by incorporating random prior knowledge is bounded by $\delta$, more specifically,*

$$\frac{\mathbb{E}[d(G_{\mathbf{T}},G_{\mathbf{X},\mathbf{W_R},\boldsymbol{\theta}})]}{|E_{\mathbf{T}}|}\le\frac{d(G_{\mathbf{T}},G_{\mathbf{X}})}{|E_{\mathbf{T}}|}+\delta\ . \tag{7}$$

Given the number of edges specified in the prior knowledge, procedures to compute $\theta$ are detailed in Algorithm S2 in Supplementary Information.

# 3 EXPERIMENTAL DESIGN AND RESULTS

We demonstrated the utility of the proposed approach using both simulation data and real biological data. In the simulation study, we tested on networks with different sizes and compared with peer methods the performance of overall network structure recovery, differential network identification and tolerance of false positives in the prior knowledge. In real data applications, we first learned the network rewiring of the cell cycle pathway of the budding yeast in response to oxidative stress, and then applied the method to study the different apoptotic signaling between recurring and non-recurring breast cancer tumors. Applications to study muscular dystrophy and transcription factor binding schemes are included in Supplementary Information.

## 3.1 Simulation study

We constructed a Gaussian Markov random field with $p=100$ nodes and 150 samples following the approach used in (Meinshausen and Bühlmann, 2006), and then randomly modified 10% of the edges to create two condition-specific networks with sparse changes. The details of simulation data generation procedure are provided in Supplementary Information. The number of edges in prior knowledge $M$ was set to be the number of common edges in the two condition-specific networks, and $\delta$ was set to 0.1.

To assess the effectiveness of prior knowledge incorporation and robustness of kDDN when false positive edges were present in prior knowledge, we examined the network inference precision and recall of the overall network and the differential network at different levels of false positive rate in the prior knowledge.

It is important to note that both false positives and false negatives in the prior knowledge here are with respect to the condition-specific ground truth from which the data are generated. Thus, false positives in prior knowledge may contribute more learning errors, false negatives will not worsen network learning performance (results shown in Supplementary Information).

Starting from prior knowledge without any false positive edges, we gradually increased the false positive rate in prior knowledge until all prior knowledge was false. At each given false positive rate in the prior knowledge, we randomly created 1,000 sets of prior knowledge, and compared the performance of kDDN in terms of precision and recall with two baselines: (1) a purely data-driven result without incorporating knowledge; and (2) a naïve baseline of knowledge incorporation by directly superimposing the prior knowledge network upon the purely data result. The results are shown in Figure 2(a) and Figure 2(b).

The dot-connected lines are averaged precision or recall and the box plot shows the first, second and third quartiles of precision or recall at each false positive rate in prior knowledge (with the ends of the whiskers extending to the lowest datum within 1.5 interquartile range of the lower quartile, and the highest datum within 1.5 interquartile range of the upper quartile).

Precision reflects the robustness to the false positive edges and efficiency of utilizing the information in prior knowledge. Figure 2(a) shows that, as expected, the false positive rate in prior knowledge has limited effect on the precision of kDDN (blue squared lines). With more false positives in the prior knowledge, the precision decreases but is still around the purely data baseline (brown circle lines) and much better than the naïve baseline (red diamond lines). While the naïve baseline suffers significantly from the false positives in prior knowledge as it indiscriminately accepts all edges in prior knowledge without considering evidence in the data. This observation corroborates the design of our method: to control the false detection incurred by the false positives in the prior knowledge. At the point where the false positive rate in the prior knowledge is 100%, the decrease of precision compared with the purely data based result is bounded within $\delta$.

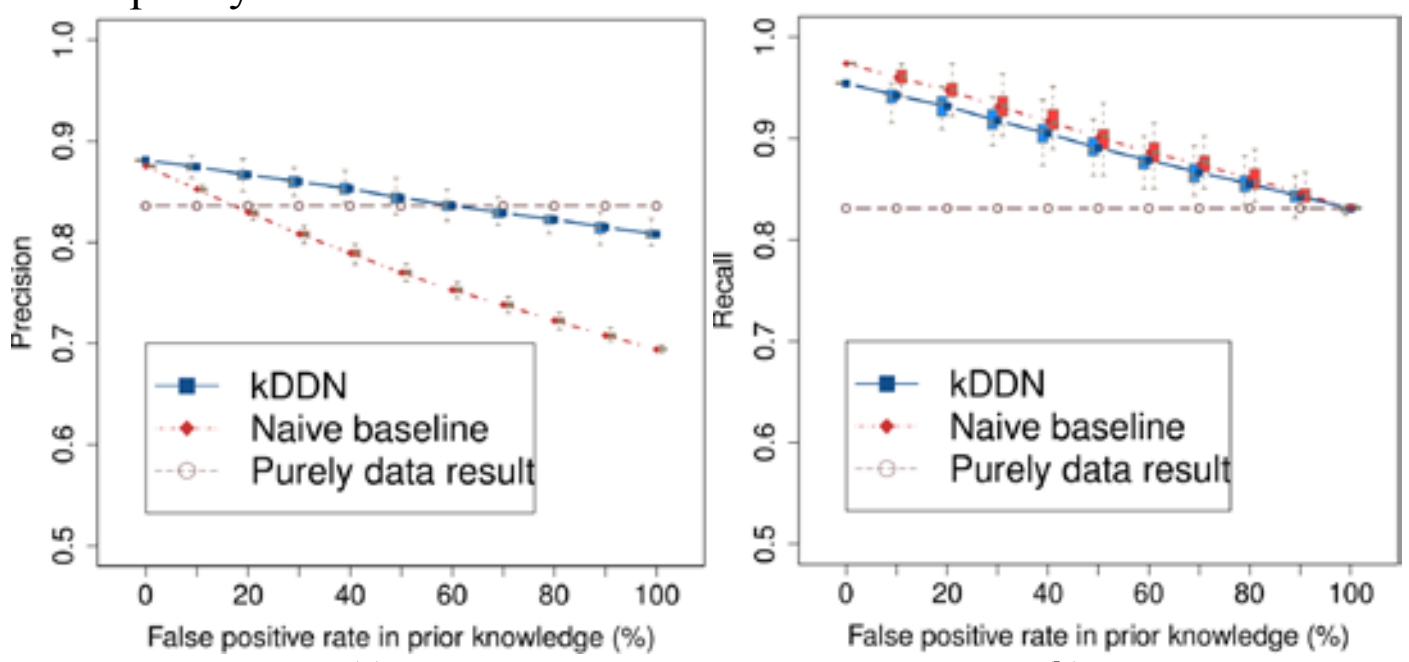

(a) (b)

**Figure 2.** The effects of false positive rate in the prior knowledge on inference precision and recall of overall network.

Recall reflects the ability of prior knowledge in helping recover missing edges. Figure 2(b) shows that when the prior knowledge is 100% false, the recall is the same as that of the purely data based result, because in this case the prior knowledge brings in no useful information for correct edge detection. When the true positive edges are included in the prior knowledge, the recall becomes higher than that of the purely data based result as more edges are correctly detected by harnessing the correct information in the prior knowledge. The naïve baseline is slightly higher in recall since it calls an edge as long as knowledge contains it, while kDDN calls an edge only when both knowledge and data evidence are present. The closeness between kDDN and naïve baseline demonstrates the high efficiency of our method in utilizing the true information in prior knowledge.

We then evaluated the effect of knowledge incorporation solely on the identification of differential network following the same protocol. The results are shown in Figure 3.

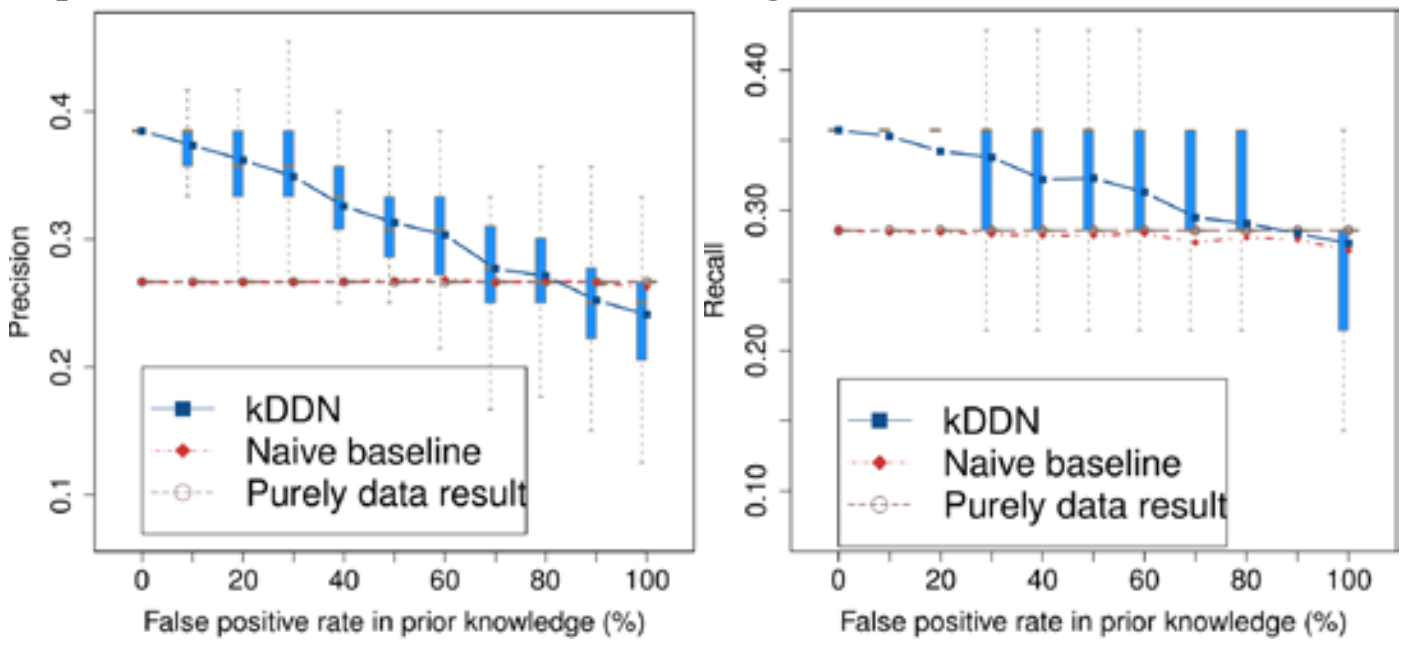

(a) (b)

**Figure 3.** The effects of false positive rate in the prior knowledge on inference precision and recall of differential network.

For differential network recovery, the naïve baseline is almost identical to purely data results because the prior knowledge seldom includes a differential edge in a large network with sparse changes. While similar advantages of kDDN apply, our method has better precision and recall, and bounds the performance degradation when knowledge is totally wrong. Unlike the naïve baseline where knowledge and data are not linked, we model the inference with knowledge and data together, so knowledge is also able to help identify differential edges.

Depending on specific conditions, false positives in prior knowledge may not distribute uniformly, but tend to aggregate more towards certain nodes. Experiments with biased knowledge distribution shown in supplementary Figures S6-S9 indicate no difference or little improvement against random knowledge, confirming that random knowledge represents the worst case and the bound according to random knowledge is sufficient.

To the best of our knowledge, our method is the first to assess the significance of differential edges. We test the type I error rate of differential edge detection of kDDN using multiple simulation data sets under the null distribution (no differential edges between the two networks) to assess if the differential edges are identified at the right significance level. If the type I error rate is either too conservative or too liberal, the *p*-value fails to reflect the actual false positive rate and we cannot control how many false positives are detected by setting a *p*-value based threshold (Chen, et al., 2011). Experiments show the average type I error rate under null distribution converges exactly to $\alpha$ under varied network sizes (Figure S12). This accuracy in *p*-value estimation gives stronger confidence in differential edge detection.

## 3.2 Performance comparisons with peer methods

We compared our joint learning method kDDN with four peer methods: 1) DDN (independent learning) (Zhang, et al., 2009), 2) csLearner (joint learning) (Roy, et al., 2011), 3) Meinshausen's method (independent learning) (Meinshausen and Bühlmann, 2006), and 4) Tesla (joint learning) (Ahmed and Xing, 2009).

csLearner can learn more than two networks but we restricted the condition to two. Meinshausen's method learns the network under single condition, and we combined the results learned under each condition to get conserved network and differential network. Tesla learns a time-evolving network, but can be adapted to two-condition learning as well. Only kDDN can assign edge-specific *p*-values to differential edges.

Parameters in kDDN are automatically inferred from data as described in Supplementary Information. For the competing methods in the comparison, we manually tested and tuned their parameters to obtain their best performance. We set DDN to detect pair-wise dependencies. The number of neighbors in csLearner is set to "4" (the ground truth value). Meinshausen's method uses the same $\lambda_1$ as inferred by kDDN as it is a special case of kDDN under one condition without prior knowledge. Tesla uses the empirically-determined optimal parameter values, since the parameter selection was not automatic but relies on user input.

To assess the impact of prior knowledge, we run kDDN under three scenarios: data-only (kDDN.dt), data plus true prior knowledge (kDDN.tk), and data plus "random" prior knowledge (kDDN.fk). Only kDDN is able to utilize prior knowledge.

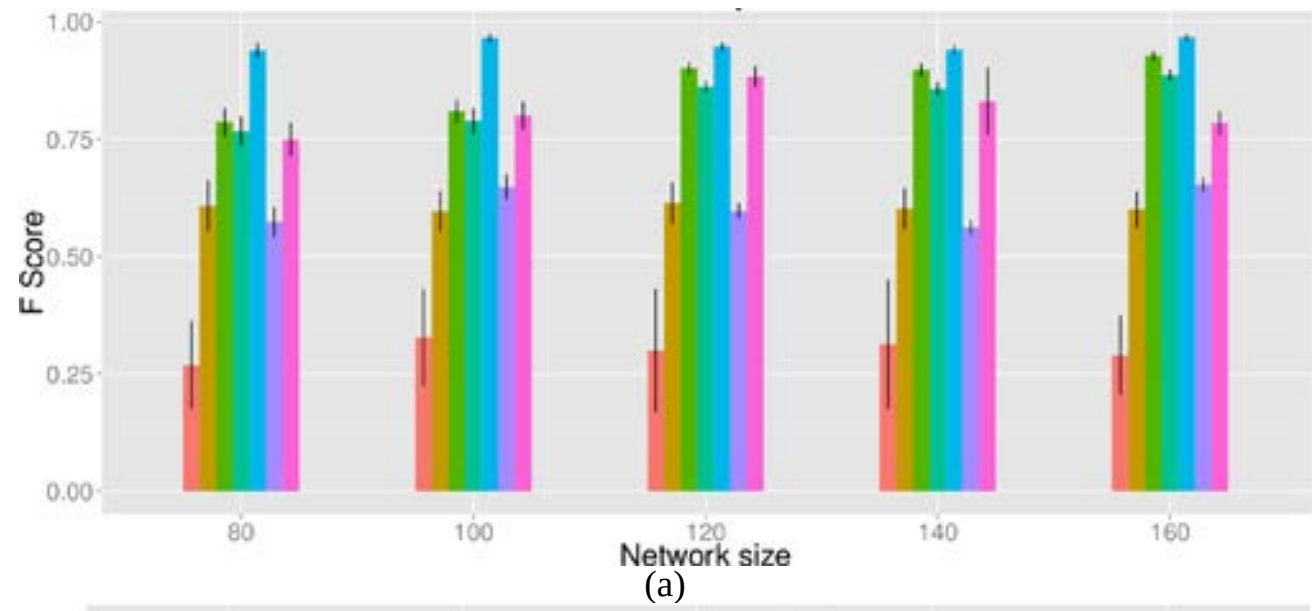

(a)

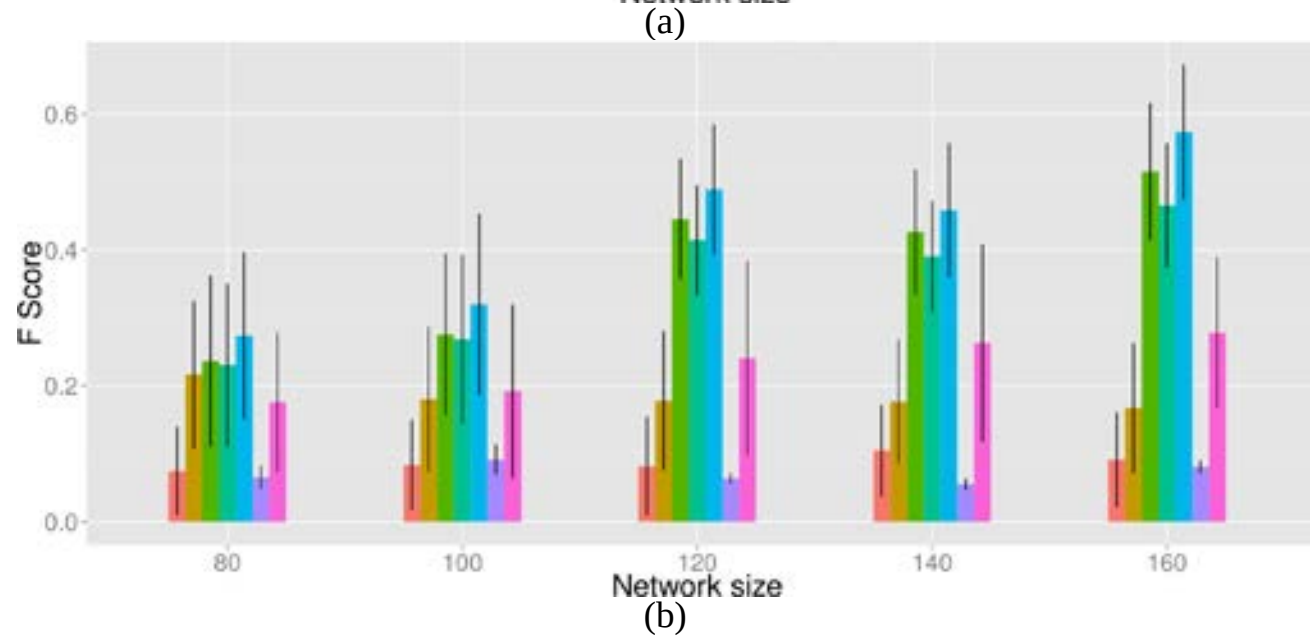

(b)

**Figure 4.** Performance comparison in F score. (a) Recovery of overall network. (b) Recovery of differential network. Legend: ■-csLearner, ■-DDN, ■-kDDN.dt, ■-kDDN.fk, ■-kDDN.tk, ■-Meinshausen, ■-Tesla.

The ground truth networks consist of 80, 100, 120, 140 and 160 nodes, respectively, and correspondingly 120, 150, 200, 200 and 240 samples. For each network size, 100 simulation datasets are generated. We evaluate the performance of inferring both overall and differential edges of the underlying networks using the F-score (harmonic mean of precision and recall, $2\frac{\text{precision*recall}}{\text{precision+recall}}$) and precision-recall averaged over all datasets under each network size.

Figure 4(a) compares the ability of recovering overall networks. We see kDDN.tk consistently outperforms all peer methods, and kDDN.dt and kDDN.fk performs comparatively to Tesla (best-performing peer method). Independent learning methods, DDN and Meinshausen's method, place third due to their inability to jointly use data.

Figure 4(b) shows the comparison of performance on recovering differential edges. Results show that kDDN consistently outperforms all peer methods under all scenarios. The fact that kDDN determines $\lambda_2$ according to the statistical significance of differential edges helps kDDN outperforms Tesla in differential edge detection. It is also clear that a single-condition method cannot find the differential edges correctly and has the worst performance.

In supplementary Figures S13 and S14, the performance of these methods is compared in precision and recall and we reach the same conclusions again.

Through these comparisons, we show that kDDN performs better than peer methods in both overall and differential network learning. High-quality knowledge further improves kDDN performance, while kDDN is robust enough to even totally random prior knowledge. Joint learning, utilization of prior knowledge and attention on statistical significance help kDDN outperform.

## 3.3 Pathway rewiring in yeast uncovers cell cycle response to oxidative stress

To test the utility of the methods in real biological study, we applied the kDDN to the public data set GSE7645. This data set used budding yeast *Saccharomyces cerevisiae* to study the genome-wide response to oxidative stress imposed by cumene hydroperoxide (CHP). Yeast cultures were grown in controlled batch conditions, in 1 L fermentors. Three replicate cultures in mid-exponential phase were exposed to 0.19 mM CHP, while three non-treated cultures were used as controls. Samples were collected at t=0 (immediately before adding CHP) and at 3,6,12,20,40,70 and 120 min after adding the oxidant. Samples were processed for RNA extraction and profiled using Affymetrix Yeast Genome S98 arrays.

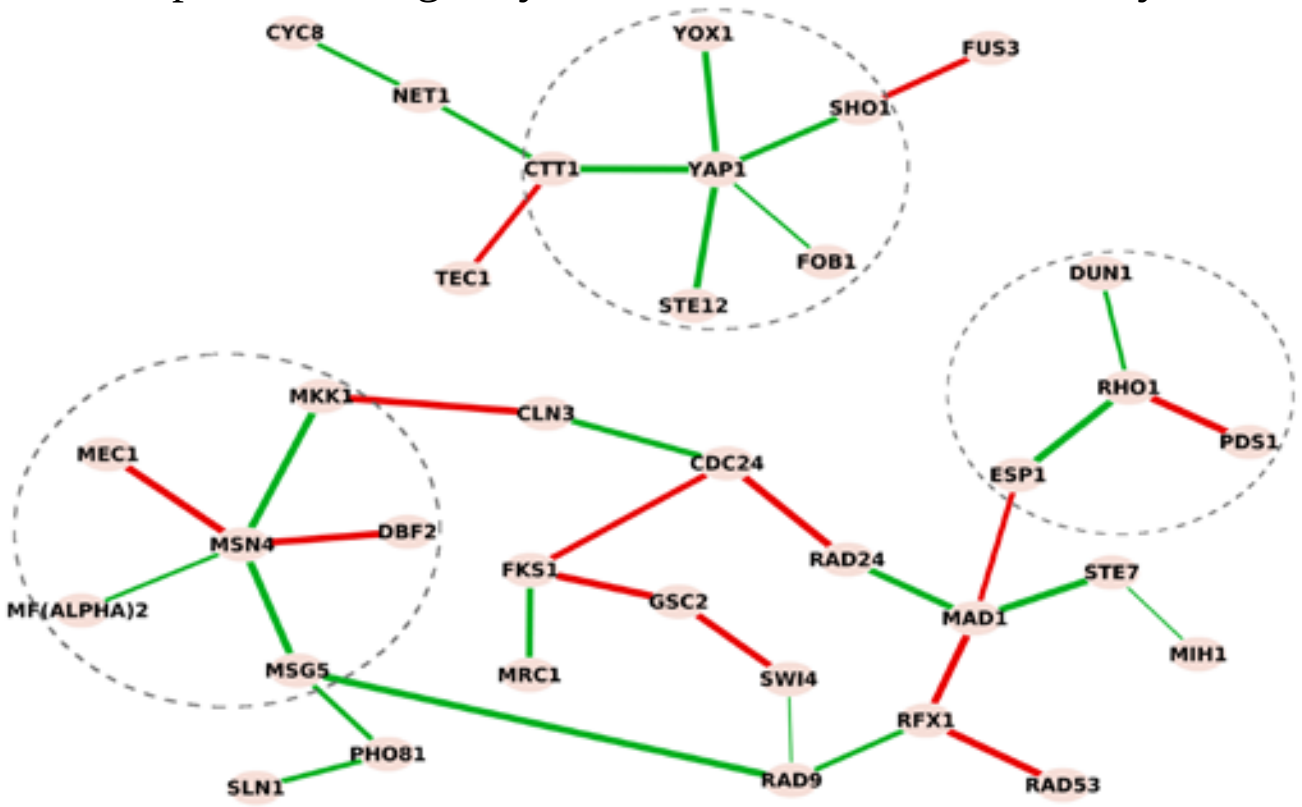

**Figure 5.** Differential dependency network in budding yeast reflects the cell cycle response to oxidative stress.

We analyzed the network changes of cell cycle related genes with structural information from the KEGG yeast pathway as prior knowledge. We added a well studied yeast oxidative stress response gene *Yap1* (Costa, et al., 2002; Ikner and Shiozaki, 2005; Jamieson, 1998; Kuge, et al., 1997) and related connections gathered from the Saccharomyces Genome Database (Cherry, et al., 2011) to the knowledge network. The learned differential network result is shown in Figure 5, in which nodes represent genes involved in the pathway rewiring, and edges show the condition-specific connections. Red edges are connections in control and green edges are connections under stress. Wider edges have higher significance.

Oxidative stress is a harmful condition in a cell, tissue, or organ, caused by an imbalance between reactive oxygen species and other oxidants and the capacity of antioxidant defense systems to remove them. The result shows that *Yap1*, *Rho1* and *Msn4* are at the center of the network response to oxidative stress; they are activated under oxidative stress and many connections surrounding them are created (green edges). *Yap1* is a major transcription factor that responds to oxidative stress (Costa, et al., 2002; Ikner and Shiozaki, 2005; Jamieson, 1998; Kuge, et al., 1997). *Msn4* is considered as a general responder to environmental stresses including heat shocks, hydrogen peroxide, hyper-osmotic shock, amino acid starvation (Causton, et al., 2001; Gasch, et al., 2000). *Rho1* is known to resist oxidative damage and facilitate cell survival (Lee, et al., 2011; Petkova, et al., 2010; Singh, 2008). The involvement of these central genes captured the dynamic response of how yeast cell sense and react to oxidative stress. The edge between *Yap1* and *Ctt1* under stress grants more confidence to the result. *Ctt1* acts as an antioxidant in response to oxidative stress (Grant, et al., 1998), and the coordination between *Yap1* and *Ctt1* in protecting cells from oxidative stress is well established (Lee, et al., 1999). This result depicted the dynamic response of yeast when exposed to oxidative stress and many predictions are supported by previous studies, which validated the effectiveness of the methods in revealing underlying mechanisms and providing potentially novel insights. These insights would be largely missed by conventional differential expression analysis as the important genes *Rho1*, *Msn4*, *Yap1* and *Ctt1* ranks 13, 20, 64 and 84 among all 86 involved genes based on *t*-test *p*-values. In a comparison with data-only results in Supplementary Information, 14 different differential edges are found.

## 3.4 Apoptosis pathway in early recurrent and non-recurrent breast cancer patient

Network rewiring analysis can be utilized to study the mechanistic differences between long-term outcomes of a disease and help find the underlying key players that cause these differences. For example, 50% of estrogen receptor positive breast cancers recur, but the mechanisms involved in causing recurrence remain unknown. Understanding of the mechanisms of breast cancer recurrence can provide critical information for early detection and prevention. We used gene expression data from a clinical study (Loi, et al., 2007) to learn differences in the apoptosis pathway in primary tumors between later recurring and non-recurring patients. We compared the pathway changes in tumors obtained from patients whose breast cancer recurred within 5 years after treatment and from patients who remained recurrence free for at least 8 years after treatment. There were 47 and 48 tumor samples in the recurring and non-recurring groups, respectively. Gene expression data were generated using Affymetrix U133A arrays. We used the apoptosis pathway from KEGG as prior knowledge.

Following the same presentation as in the yeast study, red edges are connections established in recurring patients, and green edges are connections unique to non-recurring patients. Differences in the signaling among genes in the apoptosis pathway between tumors in patients that subsequently recurred or remained cancer free are shown in Figure 6.

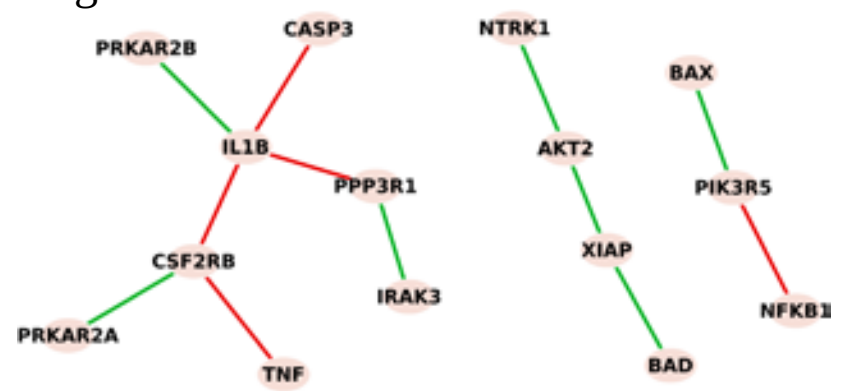

**Figure 6.** Rewiring of apoptosis pathway in breast cancer patients with/without recurrence. Red edges are connections in recurring patients, and green edges are connections in non-recurring patients.

Three inflammatory/immune response genes (*IL1B*, *NFκB* and *TNFα*) that are all linked to increased resistance to breast cancer treatment were identified in the recurring tumors. These genes formed a path to inhibit proapoptotic *CASP3* and *PPP3R1* (Su, et al., 2012), and to activate the pro-survival genes *PIK3R5* or *CSF2RB* that maintain cell survival. In contrast, green edges that were present in non-recurring tumors form paths to both anti-apoptotic *XIAP*/*AKT2* and proapoptotic *BAX* and *BAD* gene functions.

When we overlaid the differential network over the KEGG apoptosis pathway we noticed additional differences in the signaling patterns. Using the same color-coded presentation we show the learned differential network in Figure 7.

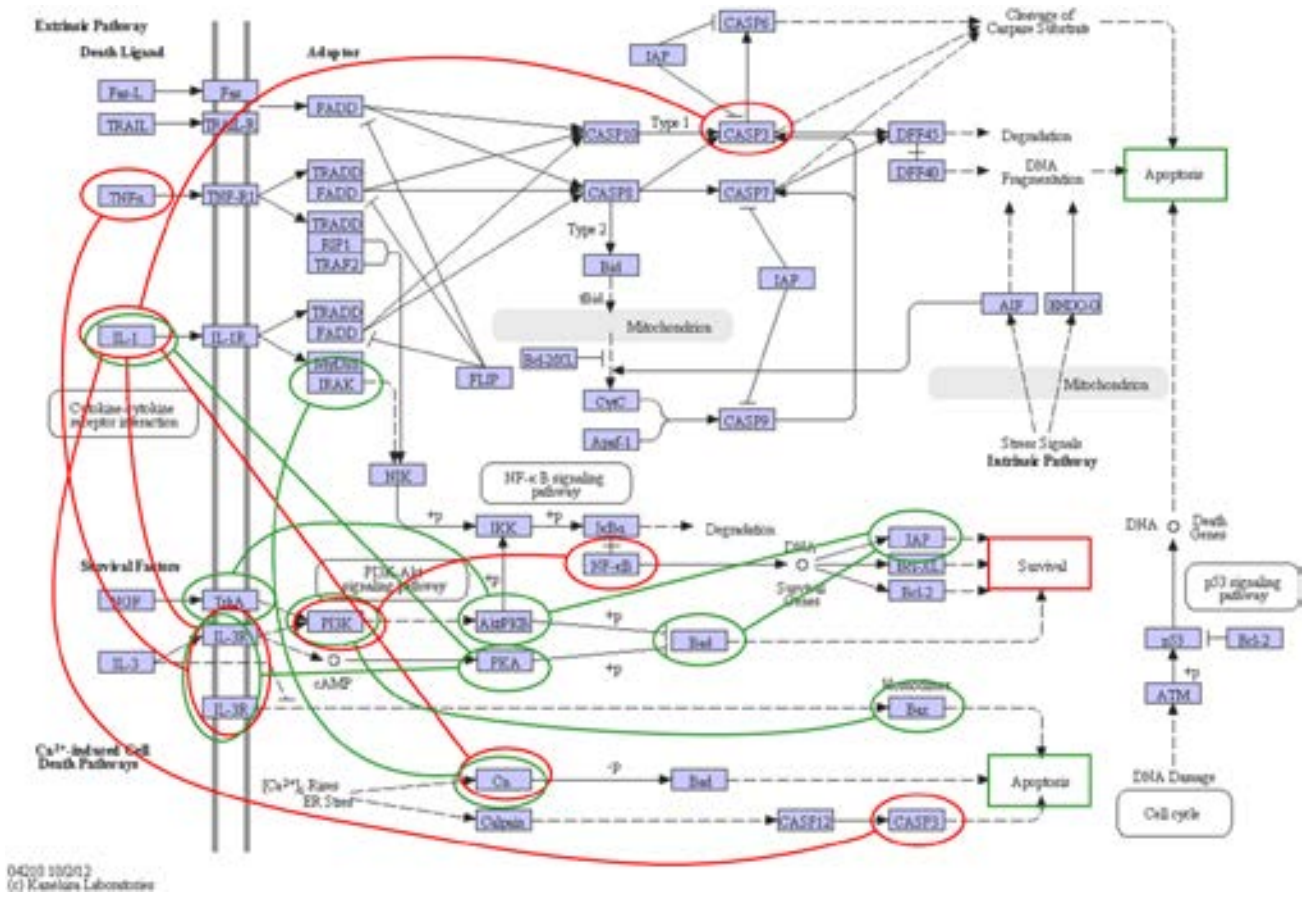

**Figure 7.** Differential network presented over KEGG apoptosis pathway.

In the recurring breast cancers (red edges), the molecular activities mainly affect the initial apoptotic signals outside the cell and within cell membrane (ligands and their receptors), while inside the cell there is no clear signaling cascade affected to determine cells fate. The only route affected within the cell is *IL1B*-induced inhibition of pro-apototic *CASP3*. In the non-recurring breast cancer, the affected network involves both signals received from activation of the membrane receptors and a cascade of signaling path inside a cell to promote apoptosis as well as survival. Since a balance between apoptosis and survival is necessary for damaged cells to be eliminated and repaired cells to survive (Murphy, et al., 2000), it is logical that both pathways would be activated concurrently.

In conclusion, the apoptosis pathway rewiring analysis identified key mechanistic signaling differences in the tumors between recurring and non-recurring patients. These differences provide a promising ground for novel hypothesis to determine factors affecting breast cancer recurrence.

## 4 DISCUSSION

To address the challenges concerning differential network inference using data-knowledge integrated approaches, we formulated the problem of learning the condition-specific network structure and topological changes as a convex optimization problem. Model regularization and prior knowledge were utilized to navigate through the vast solution space. An efficient algorithm was developed to make the solution scalable by exploring the special structure of the problem. Prior knowledge was carefully and efficiently incorporated in seeking of the balance between the prior knowledge support and data-derived evidence where our method can efficiently utilize prior knowledge in the network inference while remaining robust to false positive edges in the knowledge. The statistical significance of rewiring and desired type I error rate were assessed and validated. We evaluated the proposed method using synthetic data sets in various cases to demonstrate the effectiveness of this method in learning both common and differential networks, and the simulation results further corroborated our theoretical analysis. We then applied this approach to yeast oxidative stress data to study the cell dynamic response to this environmental stress by rewiring network structures, and the results were highly consistent with previous findings, providing meaningful biological insights into the problem. Finally, we applied the methods to breast cancer recurrence data and obtained biologically plausible results. In the future, we plan to incorporate more types of biological prior information, *e.g.*, the protein-DNA binding information in ChIP-chip data and protein-protein interaction data, and work on the improvement to utilize condition-specific prior knowledge.

## ACKNOWLEDGEMENTS

This work is supported in part by the National Institutes of Health under Grants CA160036, CA164384, NS29525, CA149147, and HL111362.